\documentclass[twocolumn,prb,preprintnumbers,amsmath,amssymb]{revtex4}
\usepackage[T1]{fontenc}
\usepackage{graphics,graphicx,amsmath}
\usepackage{dcolumn}

\begin{document}
\newcommand{\zl}[2]{$#1\:\text{#2}$}
\newcommand{\zlm}[2]{$#1\:\mu\text{#2}$}
\title{Development of a Fast Position-Sensitive Laser Beam Detector}
\date{\today}

\author{Isaac Chavez\footnote[1]{chavez@physics.utexas.edu}}
\author{Rongxin Huang}%
\author{Kevin Henderson\footnote[2]{Currently at Los Alamos National Laboratory,Los Alamos, NM 87545}}
\author{Ernst-Ludwig Florin}
\author{Mark G. Raizen}
\affiliation{%
Center for Nonlinear Dynamics and the Department of Physics, University of Texas at Austin, Austin, Texas 78712\\}%

\begin{abstract}
We report the development of a fast position-sensitive laser beam detector with a bandwidth that exceeds currently available detectors.  The detector uses a fiber-optic bundle that spatially splits the incident beam, followed by a fast balanced photo-detector. The detector is applied to the study of Brownian motion of particles on fast time scales with 1 \AA{ngstr\o{m}} spatial resolution. Future applications include the study of molecule motors, protein folding, as well as cellular processes.
\end{abstract}

\maketitle
\section{\label{sec:level1}Introduction}
Laser beam pointing position sensing finds many important applications in atomic force microscopy, spatial imaging with optical tweezers, and target acquisition\cite{1, 2, 3, 4, 5, 6}. The simplest realization of beam pointing detection is to focus the laser onto a split detector.  A commercial realization of this concept is the quadrant detector that splits the beam into four parts, providing information about the laser pointing in two orthogonal directions.  Typically, the large surface-area of these photodiodes (and hence large capacitance), limits the detector speed to only a few hundred KHz.  The quadrant detector can be reduced in size in order to improve the bandwidth, but this approach is limited by two factors:  The first is that the area of the gap between the quadrants must be kept small compared with the active photodiode area.  The second factor is that at high frequencies there will be significant cross-talk between the quadrants, and photocurrents cannot be simply subtracted as is presently done. \par

Our interest in this problem was stimulated by the study of Brownian motion of nanometer size beads held by optical tweezers where very fast response is needed.  We report here a new split detector design which makes fast detection possible.  Our detector measures motion in one transverse direction, but can be easily generalized to two and three dimensions.  Beyond the study of Brownian motion, this detector should find important scientific and technological applications.
\section{\label{sec:level0}Design}

Our guiding philosophy in developing a new device is to separate the functions of a position sensitive detector so that each can be independently optimized.  The first function is splitting of the beam and the second function is the balanced photodetection of each half.  This is realized in practice with a fiber optic waveguide that is split into halves with the light from each end focused onto a fast photodiode.  The two detectors are operated in a balanced subtraction mode, providing position information.  A schematic of our set-up is shown in Fig. \ref{fig1} and we now describe each stage in detail. \par
\subsection{\label {sec:level1} Fiber Optic Wave-Guide Splitter}
Our design uses a fiber optic waveguide splitter to spatially split an incident laser beam.  The front end of the bundle is split in half vertically, where the two halves are separated at the back end of the bundle as shown in Fig. \ref{fig2}.  The fiber splitter consists of 1000 multi-mode fibers [model MM-S105/125-22A, \zlm{105}{m} core, \zlm{125}{m} cladding, Nufern Corporation, East Granby, CT] packed into a \zl{4.4}{m} diameter front end with two back ends composed of an equal number of fibers.  The fibers are stripped of their coating to allow a higher packing efficiency which reduces loss of the incident beam.  Stripping the protective coating exposes the glass core and cladding, and heat-shrink tubing is used to bind and protect the fibers tips.  To permanently seal the fibers, low viscosity glue [type 609, Henkel Loctite Corporation, Rocky Hill, CT] is applied between the fibers.  Two aluminum rings are epoxied over the front end to help mount the bundle for cutting and polishing.  We use a diamond edge saw to cut the fibers to the same length. \par

Polishing is necessary to remove the rough cut of the diamond edge saw.  We use a series of aluminum oxide lapping films of grits ranging from \zlm{30}{m} to \zlm{0.3}{m} [model LFG,  aluminum oxide lapping films:  \zlm{30}{m}, \zlm{5}{m}, \zlm{3}{m}, and \zlm{0.3}{m}, ThorLabs Corporation, Newton, NJ]. An aluminum holder keeps the fiber bundle perpendicular to the lapping film surface to ensure the smoothest and flattest polish.  The lapping films are placed on a glass surface and polishing lubricant is added.  We start with the coarsest lapping film (\zlm{30}{m}), and stepped with various lapping film grits ending with a lapping film of \zlm{0.3}{m}.  Between each lapping film polish, the bundle is inspected under a microscope to ensure that the fibers are being polished evenly.  The front end of the bundle is split after polishing is complete.\par

Splitting the front end of the fiber optic waveguide is accomplished by mounting a flat-edge mask (razor blade) vertically on a translation stage in front of the waveguide and illuminating with a laser.  The mask blocks exactly half of the fiber waveguide so sorting can be accomplished.  At the other end, fibers that emit light are manually moved to one side while the dark fibers are moved to the other side.  Fibers that are partially covered with the knife edge (center fibers) are divided up by relative intensity, providing a gradiated splitting.  Once the sorting is complete the two back ends are bound and polished in the same fashion as the front end.  A photograph of the completed waveguide is shown in Fig. \ref{fig3}.  In the final setup, the waveguide is mounted on an x-y-z translation stage to enable alignment with respect to the incident beam.
\subsection{\label {sec:level1}Optics and Detector}
\begin{figure}
\includegraphics[width=0.48\textwidth]{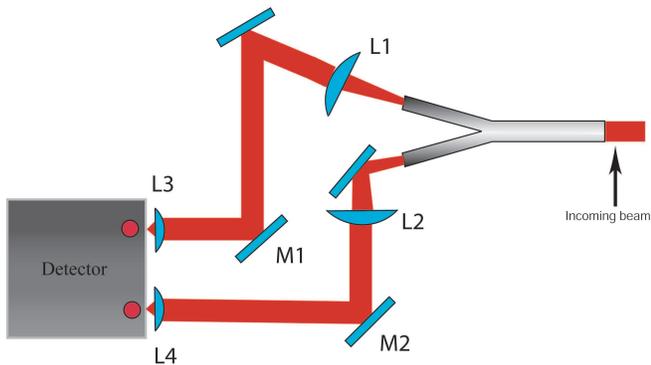}
\caption{\label{fig1} Schematic of the complete position-sensitive detector. The fiber optic waveguide splitter splits an incident beam into two equal parts.  Lenses L1 and L2 serve to collimate the light.  The light is directed with mirrors M1 and M2 and focused onto the balanced-photo detector with lenses L3 and L4.}
\end{figure}
After the waveguide splitter, the rest of the position detector consists of a balanced photo-detector and optics.  Due to the numerical aperture of the fibers (NA = 0.22), light coming out of the waveguide must be collimated and focused onto the balanced photo-detector.  In our experimental setup, we use a \zl{1064}{nm} laser with a maximum power of \zl{700}{mW} [NPRO model 126-1063-700, Lightwave Electronics (now JDSU), Milpitas, CA].\par
\begin{figure}
\includegraphics[width=0.25\textwidth]{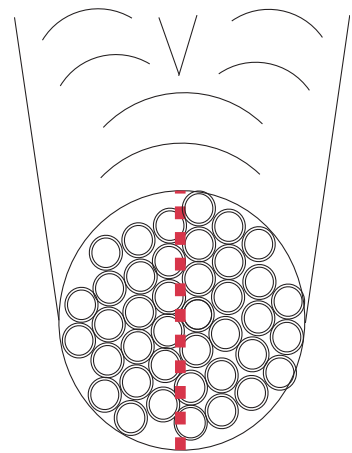}
\caption{\label{fig2} Schematic of the front end of the fiber optic waveguide splitter.  Fibers on the left side of the red dashed line are sorted into one bundle on the back end while fibers on the right side are sorted on the other end.  A laser and knife edge is used to split the waveguide into two equal parts.}
\end{figure} 
We collimate the light from the back end of each bundle using two-inch diameter, \zl{37}{mm} focal length aspheric lenses (lenses L1 and L2 in Fig. \ref{fig1}).  Two mirrors (M1 and M2) direct the collimated light towards the balanced detector.  Smaller aspheric lenses, one-inch diameter \zl{12}{mm} focal length (lenses L3 and L4), focus the light onto the detector.  Each of these lenses and the detector are mounted on translation stages for alignment and balancing.\par

We use a balanced photo-detector sensitive in the range from \zl{800}{nm} to \zl{1700}{nm} [PDB120C, ThorLabs, Newton, NJ].  The detector has a bandwidth from DC to 75 MHz and has a photodiode of diameter \zl{0.3}{mm}.  The detector is chosen for its high transimpedance gain (1.8x10$^5$ V/A) and high-frequency bandwidth.  The subtracted photocurrent provides the baseline for our measurements.  As a optically trapped particle moves within the incident beam, the fraction of light impinging on each detector changes, and the resulting voltage of the balanced detector is proportional to the particle excursion as long as the displacement remains small \cite{4}.  
\section{\label{sec:level0}Results}
\subsection{\label{sec:level1}Waveguide Splitter Efficiency}
To determine the efficiency of the waveguide we measure the fraction of light that is transmitted by the bundle.  We mount the waveguide on an x-y-z translation stage and we set up two power meters to measure the intensity coming out each back end.  A YAG \zl{1064}{nm} laser (beam waist of \zl{4}{mm}), illuminates the front end of the waveguide.  We adjust the waveguide until the beam is centered on the front end and the intensity coming out each back end is equal. The "balanced power" is determined to be $30\%$ of the initial power of the initial laser beam. The loss in intensity is due to packing efficiency, broken fibers (5 fibers were known to have broken out of 1000), and the fiber cladding. The fiber can only transmit light if the light impinges on the core of the fiber therefore light entering the cladding is lost.  A loss of 
$29\%$ is accounted for by the ratio of the cladding area to the area of the total fiber.
\subsection{\label{sec:level1}Temporal and Spatial Resolution}
To test the temporal response of our detector, we record Brownian motion of a single particle.  The experimental setup consists of a focused \zl{1064}{nm} YAG laser beam (optical tweezer\cite{7}) that creates a harmonic potential for the small bead in the center of the focus\cite{8}.  As the particle moves within this potential it changes the distribution of intensity within the laser beam.  The transmitted light is collected by a condenser lens and collimated by a \zl{25}{mm} focal length lens.  This collimated beam goes into our waveguide splitter and onto our balanced detector.  We use attenuators between the \zl{25}{mm} focal length collimating lens and the waveguide splitter to control the power impinging on our detector.  We test both polystyrene and silica beads of diameter of \zlm{0.5}{m} and \zlm{1}{m} in water.  We use a high resolution 14-Bit, DC to 100MHz high-speed digitizer made by National Instruments (model PXI-5122).  All data sets are taken with a sample rate of \zl{100}{MS/s} with a signal shot.  The data is collected on the digitizer board and transferred to the computer for analysis. \par
\begin{figure}
\includegraphics[width=0.5\textwidth]{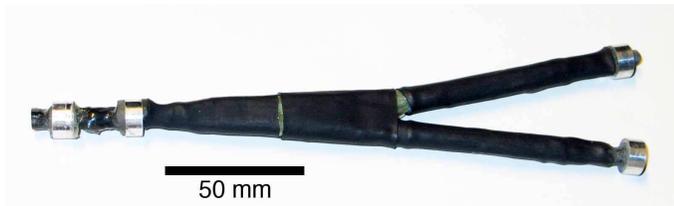}
\caption{\label{fig3} Photo of completed split fiber bundle.  The bundle is \zl{20.3}{cm} in length and the two back ends are split about \zl{5}{cm}.  The aluminum rings help with mounting for cutting and polishing and heat-shrink tubing helps protect the fibers. }
\end{figure}

\begin{figure}
\includegraphics[width=0.5\textwidth]{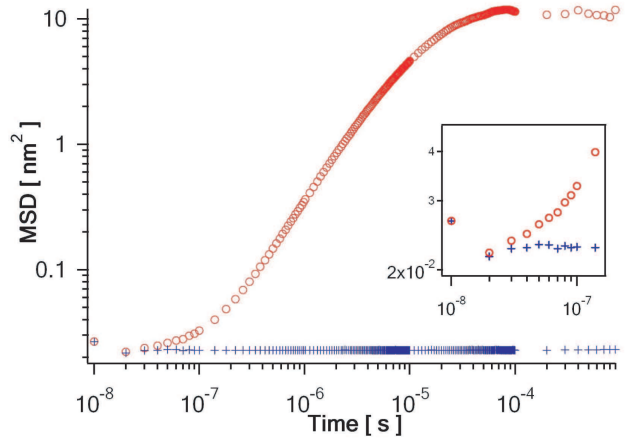}
\caption{\label{fig4}Mean square displacement of a 1mm polystyrene bead in an optical trap (red circles) with the noise floor shown (blue crosses).  The detector bandwidth is at least on the order of 30 MHz and the spatial resolution is about 1 \AA{ngstr\o{m}}.}
\end{figure}

Fig. \ref{fig4} shows the mean square displacement of a \zlm{1}{m} polystyrene bead in water.  For this curve, the y-axis shows the mean square displacement in units of {nm$^2$} and the noise floor is given to show both the temporal and spatial resolution range of the detector. The noise signal is recorded when a particle is not present within the optical trap. The range of our detector is around \zl{30}{ns} which corresponds to a mean square displacement below \zl{0.03}{nm$^2$}, providing a spatial resolution in the 1 \AA{ngstr\o{m}} range.  This result is confirmed by measuring different sized polystyrene and silica beads and we find similar spatial resolution.\par
The mean square displacement shown in Fig. \ref{fig4} agrees well with the current theory of Brownian motion on fast time scales\cite{9}. Further work is currently being conducted in this direction using our detector.

\section{\label{sec:level0}Discussion and Conclusion} 
We report the development of a position-sensitive detector by converting a balanced photo-detector into a position-sensitive detector using a fiber optic waveguide splitter. We are able to observe particle displacements on time scales shorter than \zl{100}{ns} with \AA{ngstr\o{m}} spatial resolution.  This is an important step towards the observation and exploration of Brownian motion on the fast time scale, in which the temporal and spatial resolutions are two closely coupled parameters.  A \zlm{1}{m} polystyrene particle in water moves about an average of \zl{1}{nm} in \zlm{1}{s} intervals and therefore fast detection is required not only for high-frequency events, but also to achieve high spatial resolution.  To our knowledge, no commercially available detector has the combination of frequency-bandwidth and gain. \par

The combination of temporal and spatial resolution has many applications in the study of biological processes. One example is protein folding, where fast events occur on tens of nanosecond time and \AA{ngstr\o{m}} length scale.  Another example is the study of Brownian motion in confined geometries, such as motion of particles in a living cell, which will help us understand how transport occurs within the crowded cytoplasma of living cells.  Finally, thermal motion of a particle bounded to a single molecular motor has been used to investigate its mechanical properties\cite{10}.  Using our detector, one can study motion of molecular motors on much faster time scales with higher spatial resolution which may reveal details about the conformational changes that take place when chemical energy is converted to do mechanical work.  The versatility of our design can accommodate a wide range of available balanced detectors and we expect broad range of applications beyond the discussed examples.

\section*{\label{sec:leve0}Acknowledgments}
The authors would like to thank Ed Narevicius for all his help and technical advice in the experiment. We would also like to thank the Robert A. Welch Foundation and NSF (Grants DBI-0552094, DBI-0552094, PHY-0647144) for funding this project.

\end{document}